# CeCo$_{1-x}$Fe$_x$Ge$_3$ - crystal electric field effects and fine tuning around critical substitution

P. Skokowski[1,*], K. Synoradzki[1], M. Reiffers[2], A. Dzubinska[3], Ł. Gondek[4], S. Rols[5], T. Toliński[1]

[1] *Institute of Molecular Physics, Polish Academy of Sciences, Smoluchowskiego 17, 60-179, Poznań, Poland*
[2] *Faculty of Humanities and Natural Sciences, University of Prešov, 081 16 Prešov, Slovakia*
[3] *CPM-TIP, UPJS, 040 11 Kosice, Slovakia*
[4] *AGH University of Science and Technology, Faculty of Physics and Applied Computer Science, Mickiewicza 30, 30-059 Kraków, Poland*
[5] *Institut Laue-Langevin, 6 rue Jules Horowitz, BP 156, 38042 GRENOBLE CEDEX 9, France*

Corresponding author
E-mail address: przemyslaw.skokowski@ifmpan.poznan.pl (P. Skokowski)

## Abstract

In this article we present some particularly important issues regarding the CeCo$_{1-x}$Fe$_x$Ge$_3$ alloys. Firstly, the electrical resistivity below 2 K, down to 500 mK is studied to confirm the non-Fermi-liquid behavior around the critical substitution range x~0.65. Secondly, the scheme of the crystal electric field (CEF) levels has been investigated employing methods like inelastic neutron scattering, specific heat, and magnetic susceptibility. It aims to clarify different reports on the parent CeCoGe$_3$ compound and to provide first data concerning CEF in the entire CeCo$_{1-x}$Fe$_x$Ge$_3$ series. Third, the effect of hydrogenation, especially around the quantum critical point (QCP) (x~0.65) is verified.

## 1. Introduction

We have recently carried out intensive studies of the alloys series $CeCo_{1-x}Fe_xGe_3$. This series represents a progressive transition [1] between $CeCoGe_3$, which shows three antiferromagnetic (AFM) transitions [2-5] and $CeFeGe_3$ being a heavy fermion paramagnetic compound [6]. Based on magnetometric, transport and specific heat measurements we have constructed magnetic phase diagram [7,8] and using X-ray photoelectron spectroscopy (XPS) and *ab-initio* calculations of the electronic structure we have concatenated the physical properties of these alloys with the electronic structure [9]. The latter is especially important as the transformation between the antiferromagnetic $CeCoGe_3$ and the heavy fermion, paramagnetic $CeFeGe_3$ compound is isostructural but not isoelectronic. However, for a full picture of physical properties, especially concerning the temperature dependences, a reliable knowledge of the crystal electric field (CEF) levels is useful. Due to some discrepancies in literature data [2,4] concerning the energy of the first excited doublet we analyze in the current paper the CEF scheme employing methods like specific heat, magnetic susceptibility, and inelastic neutron diffraction. Moreover, in the previous studies [7] we have provided a wide evidence of the non-Fermi liquid (NFL) behavior for x~0.65 but the experiments were carried out down to 1.9 K. In the present study we support the previous conclusions by measurements of the electrical resistivity down to temperature of 500 mK. Simultaneously, we test two methods of fine tuning of the system towards the quantum critical point (QCP), being the possibly origin of the NFL behavior: (i) the low temperature resistivity is measured at various magnetic fields and (ii) hydrogenation is performed in order to modify slightly the chemical pressure, i.e. to shift the system location on the magnetic phase diagram.

## 2. Experimental details

The samples of mass of about 6g were obtained using induction melting and the details are described in previous papers [7-9]. The inelastic neutron scattering (INS) measurements were carried out on the thermal time-of-flight (TOF) neutron spectrometer IN4 at the Institute Laue Langevin (ILL). The spectra were collected with neutron wavelength of 1.5 Å and 3 Å. The technical details are like was presented previously [10,11].

Quantum Design Physical Property Measurement System (QD-PPMS) was used to measure the magnetic susceptibility, specific heat, and electrical resistivity in the temperature range 1.9 K – 400K.

The electrical resistivity studies below 1.9 K, down to 500 mK and up to 9 T were carried out with $He^3$ refrigerator using a Q-D PPMS (in Kosice) with 4 contacts setup by AC transport method.

The hydrogenation was performed using Setaram PCT-PRO Sieverts (volumetric) system in Kraków. The powdered sample of known mass was placed in the Inconel steel reaction chamber (equipped with heater and thermocouple), which was connected to the system. At first the sample and system were flushed several times with high purity He (6N) in order to remove air. Secondly, the sample chamber was evacuated (by turbomolecular pump) at temperatures between 290-400°C for 12h to desorb spurious gases from the specimen surface. After cooling down to 30°C the dead volume of the sample container was estimated by expanding He gas from calibrated volume, which temperature was also stabilized at 30°C. Finally, the gaseous $H_2$ (6N) was introduced to the reservoir and after stabilization of the temperature and pressure the H was expanded to the sample chamber. The uptake of the hydrogen was calculated by monitoring of the pressure drop. The initial pressure applied to the sample was between 10 – 100 bar and sample temperature was raised up to 200°C in order to find proper thermodynamic conditions. It was established that for $CeCo_{0.7}Fe_{0.3}Ge_3$ at $H_2$ pressures above 50 bar and temperatures above 170°C the sample decomposes into simple hydrides. The highest hydrogen concentration (1.2 H/f.u.) was introduced at 70 bar and 30°C. The resulting sample was checked by X-ray diffraction (XRD) be means of Panalytical Empyrean diffractometer to confirm that the sample exhibits the same crystal structure as the parent compound. For second specimen $CeCo_{0.5}Fe_{0.5}Ge_3$ the higher annealing temperature before hydrogenation was applied, however even at 100 bar of $H_2$ and 200°C only 0.2 H/f.u. was introduced into sample. Once again, XRD studies revealed that the sample has the same structure as the parent compound.

### 3. Low temperature resistivity

According to our previous studies the series of alloys $CeCo_{1-x}Fe_xGe_3$ exhibits features indicating on possible QCP for x~0.65. This critical substitution has been estimated by analysis of magnetic susceptibility, electrical resistivity, thermoelectric power, and specific heat for temperatures down to 1.9 K [7]. The resistivity measurements at temperatures down to 0.5 K confirm the presence of non-Fermi liquid behavior in the vicinity of QCP.

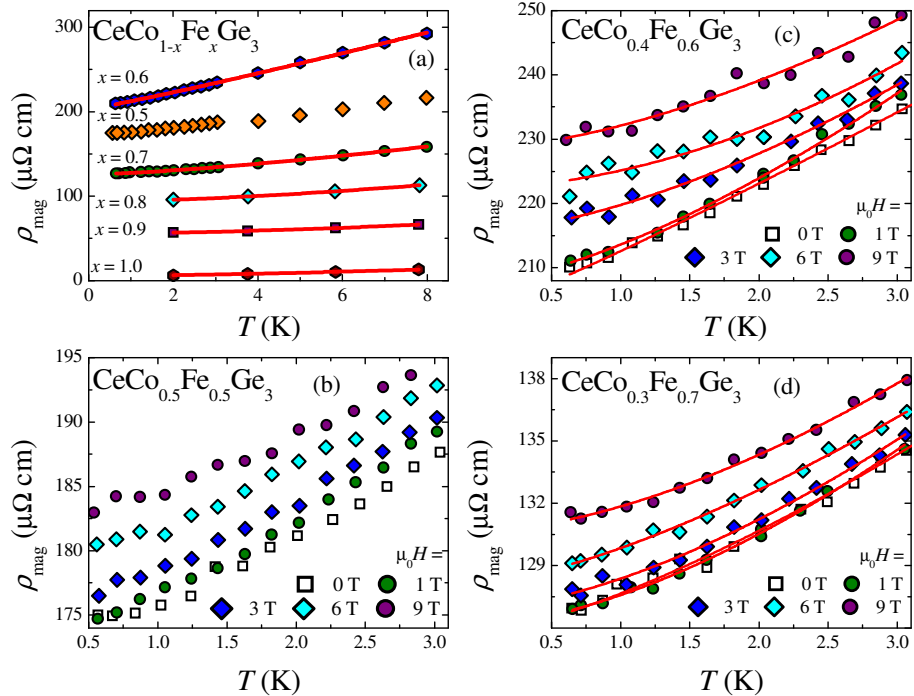

Fig. 1. Magnetic part of resistivity $\rho_{mag}$ at low temperatures for $CeCo_{1-x}Fe_xGe_3$. Results for $T >$ 1.9 K where taken from [7]. (a) $\rho_{mag}(T)$ dependence measured in zero magnetic field. Panels (b), (c), and (d) shows temperature variation of the $\rho_{mag}$ measured in several different magnetic fields for samples with x= 0.5, 0.6 and 0.7, respectively. Solid red lines represent the result of fitting the $\rho_0 + AT^\alpha$ relation to the experimental data.

**Table 1.** Parameters obtained from fitting of the formula $\rho_{mag} = \rho_0 + AT^\alpha$ to experimental data of CeCo$_{1-x}$Fe$_x$Ge$_3$.

| x (Fe) | $\rho_0$ (µΩ cm) | $\alpha$ | A (µΩ cm/K) | $\mu_0 H$ (T) | temperature range (K) | Ref. |
|---|---|---|---|---|---|---|
| 0.6 | 203(1) | 1.10(2) | 9(1) | 0 | | |
| | 208(1) | 1.5(1) | 6(1) | 1 | | |
| | 216(2) | 1.6(2) | 4(1) | 3 | 0.5 - 8 | this work |
| | 222(1) | 1.8(4) | 3(1) | 6 | | |
| | 229(2) | 1.6(3) | 4(1) | 9 | | |
| 0.7 | 125.7(2) | 1.37(3) | 1.9(1) | 0 | | |
| | 126.1(2) | 1.6(1) | 1.4(2) | 1 | | |
| | 126.9(3) | 1.6(1) | 1.4(3) | 3 | 0.5 - 8 | this work |
| | 128.2(3) | 1.4(1) | 1.7(2) | 6 | | |
| | 130.7(2) | 1.6(1) | 1.2(2) | 9 | | |
| 0.8 | 90.7(4) | 1.54(8) | 0.87(13) | 0 | 1.9 - 8 | [7] |
| 0.9 | 56.88(16) | 1.72(6) | 0.31(4) | 0 | 1.9 - 8 | [7] |
| 1.0 | 7.5(3) | 1.84(3) | 0.141(14) | 0 | 1.9 - 8 | [7] |

Fig. 1 shows magnetic part of electric resistivity $\rho_{mag}(T)$ in the range from 8 K down to 500 mK for selected samples taken in several different magnetic fields. The procedure of determining the $\rho_{mag}(T)$ with the use of La-based analogs was described in the previous paper [7]. $\rho \sim T$ is one of signatures of NFL behavior and it is clear from Fig. 1(a) that the CeCo$_{1-x}$Fe$_x$Ge$_3$ series is closest to QCP for x between 0.6 and 0.7. For x = 0.5 there is a signature of magnetic ordering at $T_1 \sim 3$ K, therefore x = 0.5 is not analyzed from the point of view of NFL state. Chemical pressure is often used as the control parameter tuning the system towards QCP, like in the above example, but, alternatively, magnetic field can be used in some cases. For the critical substitution range, i.e. for x = 0.5, 0.6, and 0.7, electrical resistivity was measured at increasing magnetic field. The $\rho_{mag}(T)$ dependences were fitted with the relation $\rho_{mag} = \rho_0 + AT^\alpha$, where for NFL the value of $\alpha$ exponent close to 1 is expected. The parameters values are collected in Table 1 and show that the critical range is expected within the x range of 0.6-0.7. For the $\rho_{mag}(T)$ dependences at magnetic fields [Fig. 1(c,d)] the exponents reveal increasing values suggesting that the system moves away from the QCP. It seems that magnetic field can be a good control parameter in the case of x = 0.5 (not analyzed due to anomaly at 3 K) but for magnetic of 3 T the anomaly seems to be removed and the dependence becomes linear.

### 4. Hydrogenation around QCP

Hydrogenation was carried out at conditions summarized in Table 2. The stable samples were chosen for further characterization by magnetic measurements. Owing to the small hydrogen dimension it can be used as a dopant without a change of the crystal structure of the host, therefore we have tested if a fine tuning of the interactions in the neighborhood of QCP is possible in this way. Figure 2(a) shows temperature dependence of the magnetic susceptibility in ZFC (zero field cooling) and FC (field cooling) mode for the hydrogen – free and hydrogenated sample with x = 0.3. The hydrogen addition nearly doubles the low temperature susceptibility values. Next step is to go towards the region of quantum critical point.

**Table 2.** Hydrogenation conditions for selected sample of $CeCo_{1-x}Fe_xGe_3$ series.

| Sample | conditions at high vacuum | pressure / concentration [H/f.u.] / comment |
|---|---|---|
| $CeCo_{0.7}Fe_{0.3}Ge_3$ | 290°C, 12h | 10 bar / 0 / up to 200°C |
| | | 25 bar / 0 / up to 200°C |
| | | 50 bar / 0.3 / decomposition at 170°C |
| | | 75 bar / 1.2 / 30°C, 16h, stable |
| $CeCo_{0.5}Fe_{0.5}Ge_3$ | 290°C, 12h | 10 bar / 0 / up to 200°C |
| | | 25 bar / 0 / up to 200°C |
| | | 75 bar / 0 / up to 200°C |
| | | 100 bar / 0.1 / up to 200°C |
| | 400°C, 12h | 10 bar / 0 / up to 200°C |
| | | 25 bar / 0 / up to 200°C |
| | | 75 bar / 0.1 / up to 200°C |
| | | 100 bar / 0.2 / 200°C, 36h, stable |

Figure 2(b) presents magnetic susceptibility for samples with x = 0.5 and x = 0.6 (FC and ZFC modes provide similar curves for these compositions) and additionally a result for x = 0.5 after hydrogenation is included. The addition of hydrogen causes dramatic change of the $\chi(T)$ dependence making it much higher and sharper and it deviates from the typical Curie-Weiss law. It is also visible that hydrogenation significantly increases the temperature independent contribution to magnetic susceptibility. The low temperature behavior of hydrogenated x = 0.5 sample seems to be closer to the case of x = 0.6. Hydrogenation appears to affect strongly the magnetic characteristics of $CeCo_{1-x}Fe_xGe_3$ but more subtle doping is necessary in further studies.

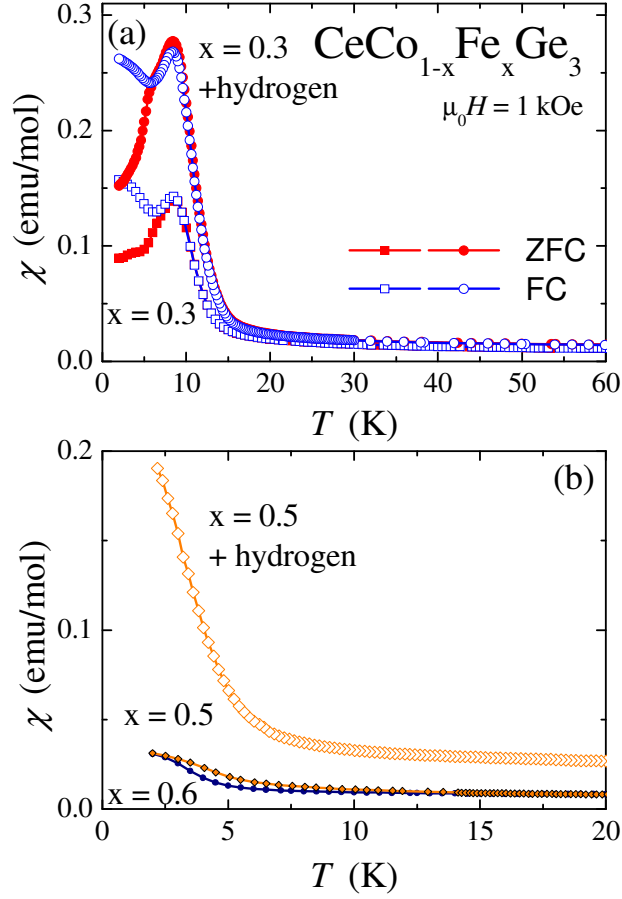

Fig. 2. (a) Magnetic susceptibility in ZFC (red) and FC (blue) mode for the hydrogen – free and hydrogenated sample of CeCo$_{0.7}$Fe$_{0.3}$Ge$_3$. (b) Magnetic susceptibility for the samples: x = 0.5, x = 0.6, and x = 0.5 after hydrogenation.

## 5. Crystal Electric Field

The interpretation of magnetic, transport and thermal properties requires knowledge of the energy levels scheme of the crystal electric field (CEF). For this purpose we performed measurements of the inelastic neutron scattering (INS) and the estimation of the CEF levels was supported by analysis of specific heat and magnetic susceptibility. The primary INS studies known from literature reveal a discrepancy in the values of the CEF splitting in CeCoGe$_3$. For single crystal Thamizhavel et al. [2] have got the crystal field parameters listed in Tab. 3 and the energy levels are $\Delta_1$ = 114 K and $\Delta_2$ = 328 K. Smidman et al. [4] for polycrystalline compound have obtained different values of the Hamiltonian parameters as is shown in Tab. 3 with the levels $\Delta_1$ = 220 K and $\Delta_2$ = 314 K. It is clear that the discrepancy concerns mainly the first excited level. Its value and degeneracy is of special importance as it decides about the low temperature properties. We have performed inelastic neutron scattering experiment for two characteristic compositions: CeCoGe$_3$ and CeCo$_{0.4}$Fe$_{0.6}$Ge$_3$, the latter being close to critical composition, i.e. to QCP. Fig. 3(a) shows the INS spectrum for CeCoGe$_3$ after subtraction of the nonmagnetic analog LaCoGe$_3$ at 50 K, i.e. in the paramagnetic state. As a result, the most probably CEF peaks seem to be located at $\Delta_1 = 220$ K and $\Delta_2 = 320$ K in agreement with Ref. [4]. The analysis is more difficult for CeCo$_{0.4}$Fe$_{0.6}$Ge$_3$, as it appeared that subtraction of the LaCo$_{0.4}$Fe$_{0.6}$Ge$_3$ spectrum from the Ce-based alloy is not effective due to small shifts in energy of the phonon peaks, even after all the standard corrections. Therefore, the best identification of the CEF excitations was possible by direct comparison of the spectra as is illustrated in Fig. 3(b). As a result, the most probably CEF peaks seem to be located at 105 K, 120 K, 220 K, and 320 K. A difference at low energy transfers is probably connected with quasi-elastic scattering for the Ce-based alloy, which may be developed by short range interactions.

**Table 3.** CEF parameters derived from INS and magnetic susceptibility in Refs. [2,4]. For the current paper the parameters are derived solely from the magnetic susceptibility.

| Sample | $B_2^0(K)$ | $B_4^0(K)$ | $B_4^4(K)$ | $\Delta_1$ (K) | $\Delta_2$ (K) | Ref. |
|---|---|---|---|---|---|---|
| CeCoGe$_3$ | 7.9 | -0.9 | 0.57 | 195 | 320 | this work |
| CeCo$_{0.4}$Fe$_{0.6}$Ge$_3$ | -2.56 | 0.28 | -2.71 | 68 | 178 | this work |
| CeCoGe$_3$ single crystal | 3 | -1 | 0 | 114 | 328 | Thamizhavel et al. [2] |
| CeCoGe$_3$ polycrystalline | -7.08 | -0.15 | 4.8 | 220 | 314 | Smidman et al. [4] |

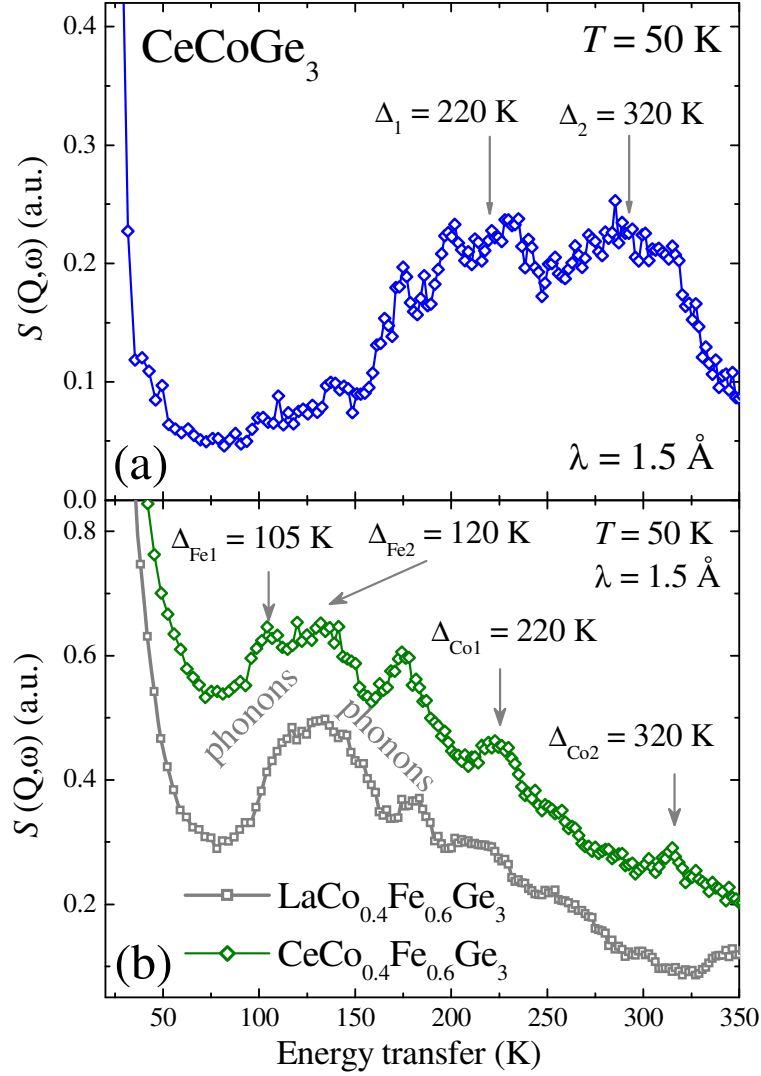

Fig. 3. INS spectra at 50 K for (a) CeCoGe$_3$ after subtraction of the LaCoGe$_3$ and (b) CeCo$_{0.4}$Fe$_{0.6}$Ge$_3$ and LaCo$_{0.4}$Fe$_{0.6}$Ge$_3$.

The presence of the excitations at 105 K and 120 K can be assigned to different CEF for Ce ions surrounded more by Fe or Co atoms. Such an interpretation is supported by specific heat measurements. It is a complementary method, which allows to estimate the CEF. Again, it is necessary to subtract $C_p(T)$ of the nonmagnetic analogue in order to get the Schottky contribution described by the formula:

$$C_{Sch}(T) = \frac{R}{T^2}\left[\frac{\sum_{i=0}^{n-1}\Delta_i^2 e^{-\Delta_i/T}}{\sum_{i=0}^{n-1} e^{-\Delta_i/T}} - \left(\frac{\sum_{i=0}^{n-1}\Delta_i\, e^{-\Delta_i/T}}{\sum_{i=0}^{n-1} e^{-\Delta_i/T}}\right)^2\right]. \qquad (1)$$

Fig. 4(a) shows the $C_p(T)$ dependence for CeCoGe$_3$ and the analogues LaCoGe$_3$, LaFeGe$_3$ and LaCo$_{0.4}$Fe$_{0.6}$Ge$_3$. This figure corroborates the observation derived from the INS experiment showing the irrelevance of LaCo$_{0.4}$Fe$_{0.6}$Ge$_3$ as the reference providing the phonon contribution. It is evident from Fig.4(a) that $C_p(T)$ of LaCo$_{0.4}$Fe$_{0.6}$Ge$_3$ shows higher values than the Ce-based alloy, therefore a subtraction procedure would lead to negative values of the Schottky contribution. However, the parent compounds, LaCoGe$_3$, LaFeGe$_3$, provide reliable values, hence we used appropriate combination of their $C_p(T)$ dependences in order to estimate the phonon part for the CeCo$_{1-x}$Fe$_x$Ge$_3$ series.

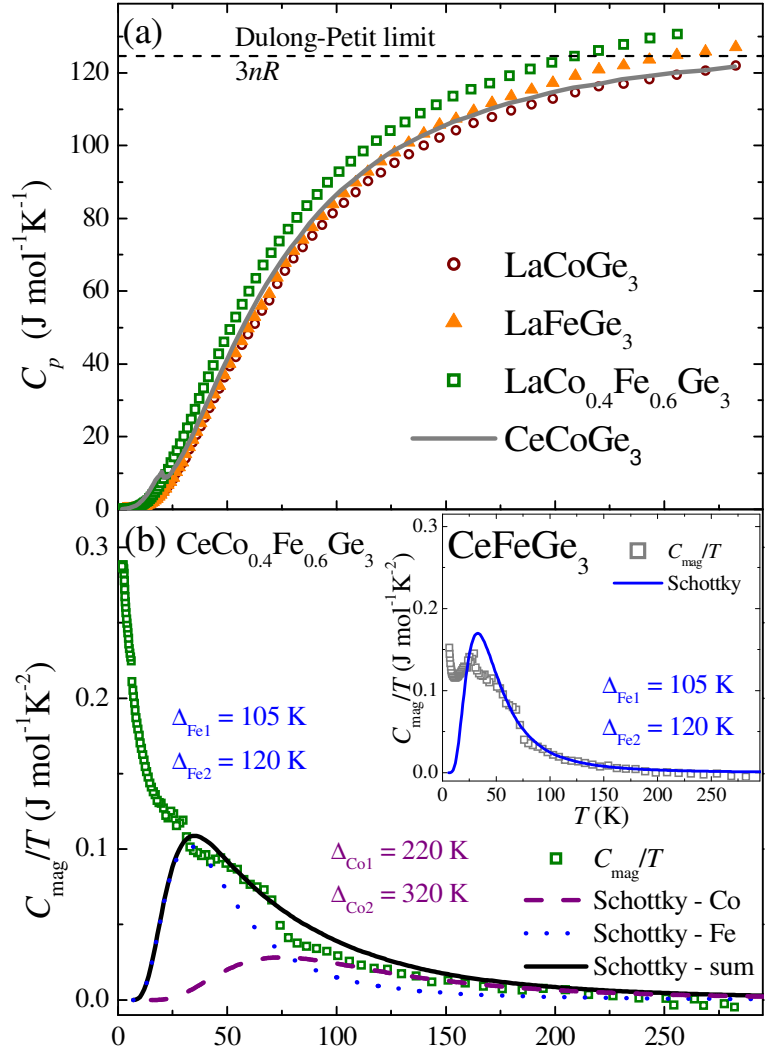

Fig. 4. Specific heat data of CeCo$_{1-x}$Fe$_x$Ge$_3$ and LaCo$_{1-x}$Fe$_x$Ge$_3$ alloys. (a) $C_p(T)$ dependence for CeCoGe$_3$ and the analogues LaCoGe$_3$, LaFeGe$_3$, LaCo$_{0.4}$Fe$_{0.6}$Ge$_3$. (b) Magnetic contribution to the specific heat for CeCo$_{0.4}$Fe$_{0.6}$Ge$_3$ and CeFeGe$_3$ (inset) analyzed with Eq. (1).

Figure 4(b) shows the magnetic contribution, obtained by the above discussed subtraction procedure, for CeCo$_{0.4}$Fe$_{0.6}$Ge$_3$. The solid lines correspond to the calculated Schottky term [Eq. (1)] using CEF energies fixed according to the results derived from INS spectra, whereas the inset shows the experimental specific heat and simulation with CEF values fixed to 105 K and 120 K.

To extend the analysis inverse magnetic susceptibility has been fitted according to formula:

$$\chi^i = \frac{N_A(g_J\mu_B)^2}{Zk_BT}\left(\sum_{\substack{n,m \\ E_n=E_m}} |\langle n|J_i|m\rangle|^2 e^{-E_n/k_BT} + 2k_BT \sum_{\substack{n,m \\ E_n \neq E_m}} \frac{|\langle n|J_i|m\rangle|^2}{E_m-E_n} e^{-E_n/k_BT}\right), \quad (2)$$

which is a sum of the Curie and Van Vleck susceptibilities. In the formula $i$ is the component of the angular momentum, $g_J$ is the Landé factor, $Z$ is the partition function, and $E_n$ represents energies of the wave functions. Finally, the fit was carried out using the formula:

$$\chi^{-1} = \left(\chi_0 + \langle \chi^i - \lambda_i \rangle\right)^{-1}, \tag{3}$$

where $\chi_0$ is the temperature independent magnetic susceptibility, $\lambda_i$ is the molecular field parameter, and triangle bracket denotes averaging of the components of the susceptibility adequately for the polycrystalline samples. The eigenenergies were derived for the CEF Hamiltonian of the form:

$$H = B_2^0 O_2^0 + B_4^0 O_4^0 + B_4^4 O_4^4, \tag{4}$$

with $O_n^m$ and $B_n^m$ being the Steven's operators and CEF parameters, respectively. The fitted inverse magnetic susceptibility is presented in Fig. 5. The key parameters are included in Tab. 3, and the other parameters are: $\chi_0 = 10\times10^{-4}$ emu/mol, $\lambda_i \sim -60$ mol/emu for each direction $i$, and a certain spread of these values is observed for x = 0.

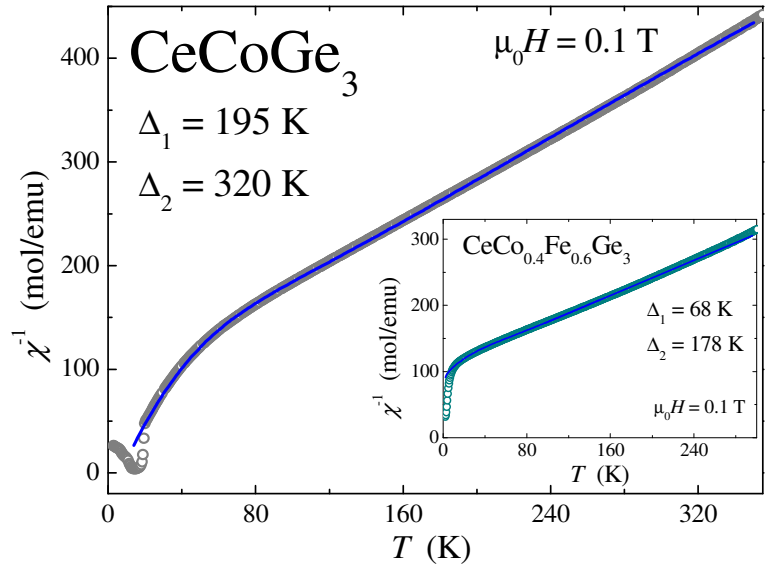

Fig. 5. Inverse magnetic susceptibility and calculation with Eq. (3) for CeCoGe$_3$ and CeCo$_{0.4}$Fe$_{0.6}$Ge$_3$ (inset).

6. Conclusions

In the current studies two main goals concerning the CeCo$_{1-x}$Fe$_x$Ge$_3$ series were achieved:
1) The non-Fermi liquid behavior for Fe content between 0.6 and 0.7 was confirmed by measurements of the electrical resistivity below 2 K. The lack of the $T^2$ dependence was observed - it appeared to be rather close to a linear one. Moreover, hydrogenation was tested as a method of fine tuning of the system towards the QCP region. We have found

2) An important challenge, in respect to literature data, is estimation of the scheme of the crystal field levels for the $CeCo_{1-x}Fe_xGe_3$ alloys. We based our analysis on the specific heat, magnetic susceptibility, and inelastic neutron scattering experiments. The overall data seem to support excitations to doublets at $\Delta_1 \sim 223$ K and $\Delta_2 \sim 315$ K for the parent $CeCoGe_3$ compound and for the $CeCo_{1-x}Fe_xGe_3$ alloy additional excitations around 105 K and 120 K seem to exist due to different local neighborhood of the Ce ions (Fe- or Co-rich).

**Acknowledgement:**

M.R. and A.D. were supported by Grants No. VEGA 1/0611/18 and No. APVV-16-0079.